\documentclass[ste,twoside]{stefano}
\usepackage{amsmath}
\usepackage{amssymb}
\usepackage{graphics}
\usepackage{rotating}
\usepackage{cite}
\usepackage{color}

\def\makeheadbox{{%
\hbox to0pt{\vbox{\baselineskip=10dd\hrule\hbox
to\hsize{\vrule\kern3pt\vbox{\kern3pt \hbox{ {\sc reply to
hep-ph/0211241} } \hbox{  {\sc
{\color{blue}{dma}}[{\color{black}{imecc}}]{\color{red}{UniCamp}}
} \hspace*{10.3cm} {\color{blue}{$\boldsymbol{\Sigma \delta
\Lambda}$}} }
\kern3pt}\hfil\kern3pt\vrule}\hrule}%
\hss}}}

\def\0{\mbox{\tiny $0$}}
\def\1{\mbox{\tiny $1$}}
\def\2{\mbox{\tiny $2$}}
\def\3{\mbox{\tiny $3$}}
\def\4{\mbox{\tiny $4$}}
\def\5{\mbox{\tiny $5$}}
\def\6{\mbox{\tiny $6$}}
\def\7{\mbox{\tiny $7$}}
\def\8{\mbox{\tiny $8$}}
\def\9{\mbox{\tiny $9$}}

\def\f14{\mbox{\tiny $\frac{1}{4}$}}
\begin{document}
%
%%%%%%%%%%%%%%%%%%%%%%%%%%%%%%%% PAPER %%%%%%%%%%%%%%%%%%%%%%%%%%%%%%%%%%%%%

\title{REPLY TO HEP-PH/0211241: ``On the extra factor of two
in the phase of neutrino oscillations''.}
%\subtitle{}

\author{
Stefano De Leo\inst{1}
%\thanks{Partially supported by the FAPESP grant 99/09008--5.}
%\and
%Gisele C. Ducati\inst{1,2}
%\thanks{Supported by a CAPES PhD fellowship.}
\and Celso C. Nishi\inst{2} \and Pietro  Rotelli\inst{3} }

\institute{
Department of Applied Mathematics, State University of Campinas\\
PO Box 6065, SP 13083-970, Campinas, Brazil\\
{\em deleo@ime.unicamp.br}
%{\em ducati@ime.unicamp.br}
%\and
%Department of Mathematics, University of Parana\\
%PO Box 19081, PR 81531-970, Curitiba, Brazil\\
%{\em ducati@mat.ufpr.br}
\and
Department of Cosmic Rays and Chronology, State University of Campinas\\
PO Box 6165, SP 13083-970, Campinas, Brazil\\
{\em ccnishi@ifi.unicamp.br}
\and
Department of Physics, INFN, University of Lecce\\
PO Box 193, 73100, Lecce, Italy\\
{\em rotelli@le.infn.it}
}

%%%%%%%%%%%%%%%%%%%%%%%%%%%%%%%%%%%%%%%%%%%%%%%%%%%%%%%%%%%%%%%%%%%%%%%%%%%
%%%%%%%%%%%% DATE ABSTRACT PACS % %%%%%%%%%%%%%%%%%%%%%%%%%%%%%%%%%%%%%%%%%

\date{{\em March, 2003}}
% Warning: Where is the date?

\abstract{ Arguments continue to appear in the literature
concerning the validity of the standard oscillation formula. We
point out some misunderstandings and try to explain in simple
terms our viewpoint.}

%%%%%%%%%%%%%%%%%%%%%%%%%%%%%%%%%%%%%%%%%%%%%%%%%%%%%%%%%%%%%%%%%%%%%%%
%%%%%%%%%%%%%%%%%%%%%%%%%%%%%%%%%%%%%%%%%%%%%%%%%%%%%%%%%%%%%%%%%%%%%%%

%%%%%%%%%%%%%%%%%%%%%%%%%%%%%%%%%%%%%%%%%%%%%%%%%%%%%%%%%%%%%%%%%%%%%%%
%%%%%%%%%%%%%%%%%%%%%%%%%%%%%%%%%%%%%%%%%%%%%%%%%%%%%%%%%%%%%%%%%%%%%%%

% Warning: No PACS code given

%02.10.Hh Rings and algebras
%02.10.Ud Linear algebra
%02.10.Yn Matrix theory

%02.30.Hq Ordinary differential equations
%02.30.Jr Partial differential equations
%02.30.Tb Operator theory

%03.65.-w Quantum mechanics
%03.65.Ca Formalism
%03.65.Ta Foundations of quantum mechanics; measurement theory

%12.15.F Quarks and lepton masses and mixing
%14.60.Pq Neutrino mass and mixing

%\offprints{~Stefano De Leo.}

\titlerunning{REPLY TO HEP-PH/0211241}

\maketitle

%%%%%%%%%%%%%%%%%%%%%%%%%%%%%%%%%%%%%%%%%%%%%%%%%%%%%%%%%%%%%%%%%%%%%%%
%%%%%%%%%%%%%%%%%%%%%%%%%%%%%%  SECTION   %%%%%%%%%%%%%%%%%%%%%%%%%%%%%
%%%%%%%%%%%%%%%%%%%%%%%%%%%%%%%%%%%%%%%%%%%%%%%%%%%%%%%%%%%%%%%%%%%%%%%

Two of the  present authors (S.D.L and P.R.) together with G. C.
Ducati  criticized a few years ago~\cite{DDR00} the derivations of
the so-called standard oscillation formula (SOF) which was then,
and remains to this date, the basis for most of the phenomenology
of neutrino masses~\cite{KAY00}. The objections then were that the
derivations of the SOF in the literature were based upon invalid
approximations~\cite{ROT01}. To be more specific the plane-wave
derivations (which are certainly the simplest) in general  ignored
the different velocities of the neutrino mass eigenstates. It is
exactly these different velocities that produce {\em slippage}
amongst the mass-eigenstate wave-packets and eventually lead to
decoherence (when oscillation ceases). An example of such a
calculation of the phase difference which makes the assumptions
$t\approx L$ and $\bar{p} \approx \bar{E}$ is
\begin{equation}
\label{sof} \Delta \, ( \, E \, t - p \, L \, )\,  = \,
 t \, \Delta E - L \, \Delta p
\,  \approx  \, t \, ( \, \Delta E - \Delta p \, ) \approx \, t \,
\, \frac{\Delta m^2}{2\, \bar{E}} ~.
\end{equation}

However,  if one allows for  the different velocities in the space
interval ($t$ constant here and $t \, \Delta v = \Delta L \neq 0
$) an extra factor of two appears in the neutrino oscillation
phase~\cite{DDR00,ROT01}
\begin{equation}
\label{f2} \Delta \, (E \, t - p \, L) \, = \,  t \, \Delta E -
\bar{L} \, \Delta p - \bar{p} \, \Delta L  \, =  \,   t \, ( \,
\Delta E - \bar{v} \, \Delta p
 - \bar{p} \, \Delta v \, )
\, \approx  \, t \,  \, \frac{\Delta m^2}{\bar{E}}~.
\end{equation}
In ref.~\cite{DDR00}, we showed that to obtain rigorously the SOF
 one needed the assumption of equal velocities. At
this point we committed  perhaps an
 ingenuity by praising the {\em aesthetic} value of equal velocities
 and this has labelled us in the eyes of some as the proponents of this
hypothesis. Some have even claimed that we believe both in an
extra factor of two and in equal velocities, notwithstanding the
fact that they are in clear contradiction.

\begin{quote}
{\sf Independently and later the equal velocity scenario was
suggested in Ref.$^2$}(ref.~\cite{DDR00} in this paper). {\sf The
authors of Ref.$^2$ consider this scenario as ``aesthetically the
most pleasing''. They proclaimed it as their ``preferred choice''
in particular because it leads to the frequency of neutrino
oscillations twice as large as the standard one} -
ref.~\cite{OT00},
\end{quote}

\noindent see also

\begin{quote}
{\sf ... the scenario of equal velocities of two mass eigenstates
is preferred in ref.[1]} (ref.~\cite{DNR02} in this paper) {\sf to
that of equal energies ...} - ref.~\cite{OST02}.
\end{quote}

In a recent work on wave packets~\cite{DNR02}, we identified the
source of the extra factor of two in the plane-wave formalism. It
is a consequence of the implicit assumption that the flavour
eigenstate  $\boldsymbol{\nu_{\alpha}}$ is identical,  including
hence its phase, at all points in the creation process,
\[ \boldsymbol{\nu_{\alpha}}=
\cos \theta \, \boldsymbol{\nu_1} + \sin \theta \,
\boldsymbol{\nu_2}~. \] This may seem very reasonable, but it is
not natural in the wave-packet formalism. In fact, within the
wave-packet formalism, the flavor eigenstate is {\em not} unique
at all points of creation. Each point is associated with an
appropriate $x$-dependent phase. For example, for gaussian wave
packets with spread $a_n$, in the $\Delta a = \Delta p =0$
scenario (with instantaneous creation)
\[ \psi(x,0) \, \boldsymbol{\nu_{\alpha}} = \left(
\frac{2}{\pi \, a^2} \right)^{\frac{1}{4}} \, \exp \left[ \, - \,
\frac{x^2}{a^2} \, \right] \, \, \exp [\, i \, p_0 \, x \, ] \, \,
\left[ \, \cos \theta \, \boldsymbol{\nu_1} + \sin \theta \,
\boldsymbol{\nu_2} \, \right] ~. \] Thus, the flavour state at
different points
 are characterized  by an
$x$-dependent phase, specifically by the plane-wave factor:
\[ \mbox{exp} \left[ \,
i \, p_0 \,  x \,  \right]~. \] In the case of different
velocities of the mass-eigenstates, interference occurs between
wave packet components corresponding to different initial wave
packet points. Thus, the final overlapping interference points
carry with them what we call an initial phase difference. This
initial phase difference compensates for the term $\bar{p} \,
\Delta L$ in Eq.(\ref{f2}), and hence eliminates the extra factor
of two, giving  the standard oscillation phase (see Section III of
ref.\cite{DNR02} for a detailed discussion).

In ref.~\cite{DNR02}, we concluded that within the wave packet
formalism the standard oscillation formula is  not only exact in
the case of equal velocities (no slippage) but also a good
approximation in all cases in which {\em minimal slippage} occurs
between the mass wave packets. Also in ref.~\cite{DNR02}, we
confronted a long standing diatribe in the literature between
equal momentum advocates and equal energy advocates of which Okun
is the most fervent proponent. The $\Delta p = 0$ hypothesis has a
mathematical "advantage", it allows one (in the wave packet
formalism) to create at a given instant
 a flavour eigenstate wave packet over an extended region.
 Flavour eigenstate creation is the starting point in oscillation
 phenomena (for kaons strong
hypercharge plays the role that lepton flavour plays for
neutrinos). {\em Unfortunately there is no physical frame in which
$\Delta p = 0$}, as can easily be shown~\cite{DNR02}. In the
physical cases in which $\Delta p \neq 0$ it is by no means
trivial to create a pure flavour eigenstate wave function. In
fact, one {\em must} allow for creation times which depend upon
the creation point, so that at no fixed instant will we have a
flavour eigenstate at all points (the "other" part of the wave
packet having evolved).

     The $\Delta E = 0$ frames do exist and so are a legitimate
     choice of frame, even if they don't happen to coincide with the
     laboratory frame in any of the
experiments, as shown by simple kinematics. As an aside, without
implying {\em any} preference, we note that only the equal
velocities case $\Delta v = 0$ is frame independent. This is a
consequence of the fact that the Lorentz transformation of
velocities is mass independent. If decoherence
 never occurs for one observer it never occurs for any observer.
 The $\Delta E = 0$ case is a choice of frame, and since oscillation
 measurement must be frame
independent, we see no reason why calculations are not made in a
manifestly Lorentz invariant manner i.e. in an arbitrary frame.

    Contrary to another of the criticisms of Okun et. al.~\cite{OST02},
     we have never assumed
interference between wave packets at different space-time points.
We have always assumed and stated that the measurement process is
made at
 a single space-time point (an idealization). At most,
 the theoretician will
have to average over the mass eigenstate wave-functions. However,
the same
 cannot be said about the creation process. The wave function is
 extended in space.
Indeed the use of a plane-wave (which is a four momentum
eigenstate) implicitly assumes a sufficient spatial extension of
the wave function to permit one to ignore the Heisenberg momentum
uncertainties ($\delta p = 0$). As for the time needed for the
creation of a wave packet, this also exists in general. In fact
even if there existed a frame in which creation where
instantaneous, another observer would "see" a finite time for
creation. This is a direct consequence of Lorentz transformations
for non-point-like entities. Hence the origin of the appearance of
multiple times and distances in our papers. Obviously, with
different velocities it is impossible to
 use a single distance-interval  and time-interval.
 Furthermore, to create a
pure flavour eigenstate we are obliged, in general, to use both
multiple distances and times. There is in this no contradiction to
quantum mechanics.

In the figure, we illustrate this in pictorial form. The two sets
of lines represent parts of a wave function for different mass
eigenstates. The vertically separation is only for design
purposes. They must be imagined initially overlapping. The cross
on the axis represents the measuring instrument in the laboratory.
The slippage of wave packets leads, at the time of measurement, to
the situation shown on the RHS where horizontal slippage has
occurred. Even assuming a common time of creation $t=0$ and of
measurement $t=T$, it is obvious that there are two different
spatial intervals $L_{\1}$ and $L_{\2}$ as displayed in the
figure. There is no sense in a common "spatial velocity" $v_s$ in
contradiction with different particle velocities $v_{\1}$ and
$v_{\2}$, as considered by Okun et al.~\cite{OST02} in their
appendix (item
2).\\

{\huge
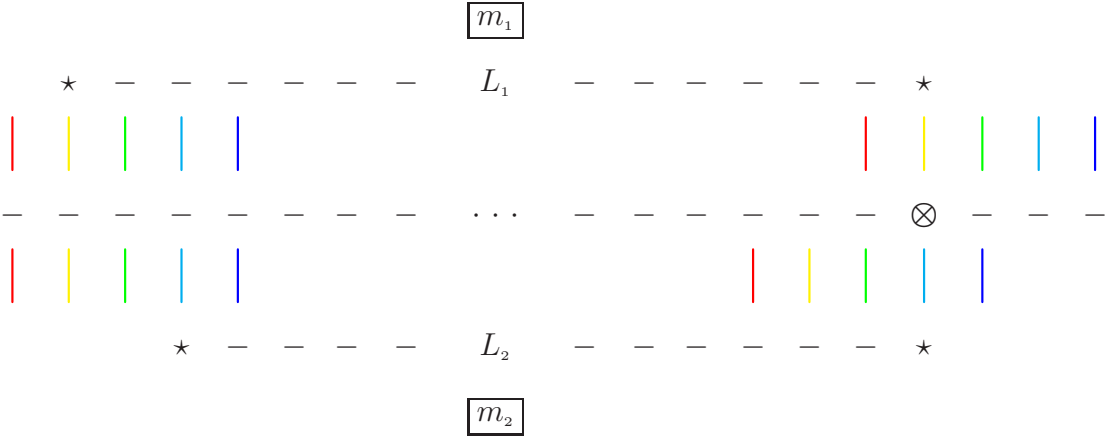
\begin{center}
\begin{tabular}{ccccccccccccccc}
  &  & & & &  &{\large \fbox{$m_{\1}$}} & & & &
&
 &   & & \\
& {\large $\star$} & {\large $-$} & {\large $-$}& {\large $-$}&
{\large $-~~-~~-~$} &{\large $L_{\1}$} &{\large $~-~~-~~-$} &
{\large $-$}& {\large $-$}& {\large $-$} & {\large $\star$}
 &   & & \\
{\color{red} $|$} & {\color{yellow} $|$}  & {\color{green} $|$} &
{\color{cyan} $|$} & {\color{blue} $|$}  & & & & & & {\color{red}
$|$} & {\color{yellow} $|$}  & {\color{green} $|$}  &
{\color{cyan} $|$} & {\color{blue} $|$}\\ {\large $-$} & {\large
$-$} & {\large $-$} & {\large $-$}& {\large $-$}& {\large
$-~~-~~-~$} &{\large $\cdot$ $\cdot$ $\cdot$} &{\large $~-~~-~~-$}
& {\large $-$}& {\large $-$}& {\large $-$} & {\large
$\boldsymbol{\otimes}$}
 & {\large $-$}  & {\large $-$}& {\large $-$}\\
{\color{red} $|$} & {\color{yellow} $|$}  & {\color{green} $|$}  &
{\color{cyan} $|$} & {\color{blue} $|$}  & & & & {\color{red} $|$}
& {\color{yellow} $|$}  & {\color{green} $|$}  & {\color{cyan}
$|$} & {\color{blue} $|$} & &\\ &  &  & {\large $\star$}& {\large
$-$}& {\large $-~~-~~-~$} &{\large $L_{\2}$} &{\large $~-~~-~~-$}
& {\large $-$}& {\large $-$}& {\large $-$} & {\large $\star$}
 &   & & \\
  &  & & & &  &{\large \fbox{$m_{\2}$}}  & & & &  &
 &   & & \\
\end{tabular}
\end{center}
}

\vspace*{.5cm}

    We have also emphasised in our preprint~\cite{DNR02} that
    the oscillation phase, and hence
oscillation formula, should depend upon the details of the
wave-packet shape and dimensions, things about which we have
little information. Again only the equal velocity case stands out
as an exception to this. This means that a single oscillation
formula will not be valid for all experiments. This should be
remembered if inconsistencies with the SOF are encountered before
invoking more exotic solutions (such as sterile neutrinos). We
believe the SOF is  a good approximation in the case of minimal
slippage between the wave-functions i.e. when
\[ t \, \Delta v (\approx L \, \Delta v)  \ll a~~~~~\mbox{[where $a$
is the wave spread]}.\]
 Otherwise one uses the SOF only on
faith.

Finally, with respect to the criticism that in a discussion about
pion decay into muon and neutrino we have adopted a mixed flavour
neutrino,

\begin{quote}
{\sf Another erroneous statement of [1]} (ref.~\cite{DNR02} in
this paper) {\sf is that in the decay $\pi \rightarrow \mu \nu$,
the $\nu$ denotes a mixture of $\nu_\mu$ and $\nu_e$} -
ref.~\cite{OST02},
\end{quote}

\noindent this is simply not true. It is an incredible criticism
since  the major part of our article~\cite{DNR02} is  {\em
devoted} to the question of guaranteeing pure flavour creation.

\end{document}